\documentclass[]{llncs}

\usepackage[dvipdfmx]{graphicx}
\usepackage{amsmath}
\usepackage{amssymb}
\usepackage{url}

\begin{document}

\title{The Complexity of (List) Edge-Coloring Reconfiguration Problem\thanks{This work is partially supported by JSPS KAKENHI Grant Numbers JP26730001 (A.~Suzuki), JP15H00849 and JP16K00004 (T.~Ito), and JP16K00003 (X.~Zhou).}}
 	\author{
 		Hiroki Osawa\inst{1} \and Akira Suzuki\inst{1,2} \and Takehiro Ito\inst{1,2} \and Xiao Zhou\inst{1}
	}
	\institute{Graduate School of Information Sciences, Tohoku University,\\
		Aoba-yama 6-6-05, Aoba-ku, Sendai, 980-8579, Japan. 
		\and
		CREST, JST, 4-1-8 Honcho, Kawaguchi, Saitama, 332-0012, Japan.\\
		\email{hiroki.osawa.r3@dc.tohoku.ac.jp \\ \{a.suzuki, takehiro, zhou\}@ecei.tohoku.ac.jp}
	}
\maketitle

\begin{abstract}
	Let $G$ be a graph such that each edge has its list of available colors, and assume that each list is a subset of the common set consisting of $k$ colors. 
	Suppose that we are given two list edge-colorings $f_0$ and $f_r$ of $G$, and asked whether there exists a sequence of list edge-colorings of $G$ between $f_0$ and $f_r$ such that each list edge-coloring can be obtained from the previous one by changing a color assignment of exactly one edge.
	This problem is known to be PSPACE-complete for every integer $k \ge 6$ and planar graphs of maximum degree three, but any complexity hardness was unknown for the non-list variant. 
	In this paper, we first improve the known result by proving that, for every integer $k \ge 4$, the problem remains PSPACE-complete even if an input graph is planar, bounded bandwidth, and of maximum degree three.
	We then give the first complexity hardness result for the non-list variant: 
for every integer $k \ge 5$, we prove that the non-list variant is PSPACE-complete even if an input graph is planar, of bandwidth linear in $k$, and of maximum degree $k$.
\end{abstract}

	\begin{figure}[t]
		\begin{center}
			\includegraphics[width=0.8\textwidth]{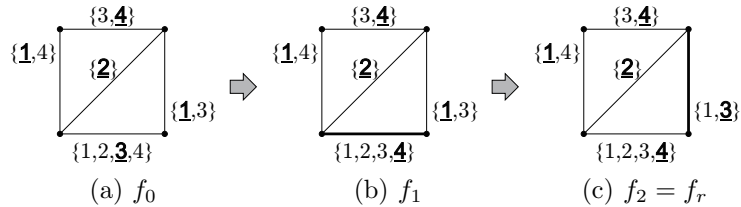}
			\\
			\hspace{5mm} (a) $f_0$ \hspace{25mm}(b) $f_1$ \hspace{20mm}(c) $f_2 =f_r$
		\end{center}
		\vspace{-1em}
		\caption{A sequence of list edge-colorings of the same graph with the same list.}
		\label{fig:LECexample}
	\end{figure}

\section{Introduction}\label{sec:01}
	Recently, reconfiguration problems~\cite{IDHPSUU11} have been intensively studied in the field of theoretical computer science.
	The problem arises when we wish to find a step-by-step transformation
	between two feasible solutions of a combinatorial (search) problem such that all intermediate results are also
	feasible and each step conforms to a fixed reconfiguration rule, that is, an adjacency relation
	defined on feasible solutions of the original search problem.
	(See, e.g., the survey~\cite{H13} and references in~\cite{DDFHIOOUY15,IOO15}.)

	\subsection{Our problem}

	In this paper, we study the reconfiguration problem for (list) edge-colorings of a graph~\cite{IKD12,IKZ12}.
	Let $C = \{ 1, 2, \ldots, k \}$ be the set of $k$ colors.
	A (proper) {\em edge-coloring} of a graph $G = (V,E)$ is a mapping $f : E \to C$ such that $f(e) \neq f(e^\prime)$ holds for every two adjacent edges $e, e^\prime \in E$.
	In {\em list edge-coloring}, each edge $e \in E$ has a set $L(e) \subseteq C$ of colors, called the \textit{list} of $e$.
	Then, an edge-coloring $f$ of $G$ is called a {\em list edge-coloring} of $G$ if $f(e) \in L(e)$ holds for every edge $e \in E$.
	Figure~\ref{fig:LECexample} illustrates three list edge-colorings of the same graph $G$ with the same list $L$; 
the list of each edge is attached to the edge, and the color assigned to each edge is written in bold with an underline.
	Clearly, an (ordinary) edge-coloring of $G$ is a list edge-coloring of $G$ for which $L(e) = C$ holds for every edge $e$ of $G$, and hence list edge-coloring is a generalization of edge-coloring. 

	Ito et al.~\cite{IKZ12} introduced an adjacency relation defined on list edge-colorings of a graph, and define the {\sc list edge-coloring reconfiguration} problem, as follows: 
	Suppose that we are given two list edge-colorings of a graph $G$ (e.g., the leftmost
	and rightmost ones in \figurename~\ref{fig:LECexample}), and we are asked whether we can transform one into the
	other via list edge-colorings of $G$ such that each differs from the previous one in only one
	edge color assignment;
such a sequence of list edge-colorings is called a {\em reconfiguration sequence}.
	We call this problem the {\sc list edge-coloring reconfiguration} problem.
	For the particular instance of \figurename~\ref{fig:LECexample}, the answer is ``yes'' as illustrated in the figure,
	where the edge whose color assignment was changed from the previous one is depicted by a thick line.
	
	For convenience, we call the problem simply the {\em non-list variant} (formally, {\sc edge-coloring reconfiguration}) if $L(e) = C$ holds for every edge $e$ of a given graph.

	\subsection{Known and related results}
	
	Despite recent intensive studies on reconfiguration problems (in particular, for graph colorings~\cite{BB13,BJLPP14,BC09,BMNR14,CH11,HIZ15,IKD12,IKZ12,JKKPP,W14}), as far as we know, only one complexity result is known for {\sc list edge-coloring reconfiguration}. 
	Ito et al.~\cite{IKD12} proved that {\sc list edge-coloring reconfiguration} is PSPACE-complete even when restricted to $k = 6$ and planar graphs of maximum degree three. 
(Since the list of each edge is given as an input, this result implies that the problem is PSPACE-complete for every integer $k \ge 6$.)
	They also gave a sufficient condition for which any two list edge-colorings of a tree can be transformed into each other, which was improved by~\cite{IKZ12};
but these sufficient conditions do not clarify the complexity status for trees, and indeed it remains open. 

	As a related problem, the {\sc (list) vertex-coloring reconfiguration} problem has been studied intensively. 
({\sc list vertex-coloring reconfiguration} and its non-list variant are defined analogously.)
	Bonsma and Cereceda~\cite{BC09} proved that {\sc vertex-coloring reconfiguration} is PSPACE-complete for every integer $k \ge 4$. 
	On the other hand, Cereceda et al.~\cite{CH11} proved that both {\sc list vertex-coloring reconfiguration} and its non-list variant are solvable in polynomial time for any graph if $k \le 3$. 
	Thus, the complexity status of {\sc vertex-coloring reconfiguration} is analyzed sharply with respect to the number $k$ of colors. 

 	Edge-coloring in a graph $G$ can be reduced to vertex-coloring in the line graph of $G$. 
 	By this reduction, we can solve {\sc list edge-coloring reconfiguration} for any graph if $k \le 3$. 
	However, the reduction does not work the other way, and hence this reduction does not yield any computational hardness result.
	Indeed, the complexity of {\sc edge-coloring reconfiguration} was an open question proposed by \cite{IKD12}. 

	\subsection{Our contribution}
	In this paper, we precisely analyze the complexity of {\sc (list) edge-coloring reconfiguration};
in particular, we give the first complexity result for the non-list variant. 

	We first improve the known result for {\sc list edge-coloring reconfiguration} by proving that, for every integer $k \ge 4$, the problem remains PSPACE-complete even if an input graph is planar, bounded bandwidth, and of maximum degree three.
	Recall that the problem is solvable in polynomial time if $k \le 3$, and hence our result gives a sharp analysis of the complexity status with respect to the number $k$ of colors.
	We then give the first complexity result for the non-list variant: 
{\sc edge-coloring reconfiguration} is PSPACE-complete for every integer $k \ge 5$ and planar graphs whose maximum degrees are $k$ and bandwidths are linear in $k$. 

	We here explain our main technical contribution roughly.
(All the details will be given later.)  
	Both our results can be obtained by constructing polynomial-time reductions from {\sc Nondeterministic Constraint Logic} (NCL, for short), introduced by Hearn and Demaine~\cite{HD05}.
	This problem is often used to prove the computational hardness of puzzles and games, because a reduction from this problem requires to construct only two types of gadgets, called {\sc and} and {\sc or} gadgets. 
	However, there is another difficulty for our problems, that is, how the gadgets communicate with each other. 
	We handle this difficulty by introducing the ``neutral orientation'' to NCL. 
	In addition, our {\sc and}/{\sc or} gadgets are very complicated, and hence for showing the correctness of our reductions, we clarify the sufficient conditions (which we call ``internally connected'' and ``external adjacency'') so that the gadgets correctly work. 
	We show that our gadgets indeed satisfy these conditions by a computer search.

\section{Nondeterministic Constraint Logic}\label{sec:02}

		In this section, we define the {\sc nondeterministic constraint logic} problem~\cite{HD05}.
		
		\begin{figure}[t]
			\begin{center}
				\includegraphics[width=0.8\textwidth]{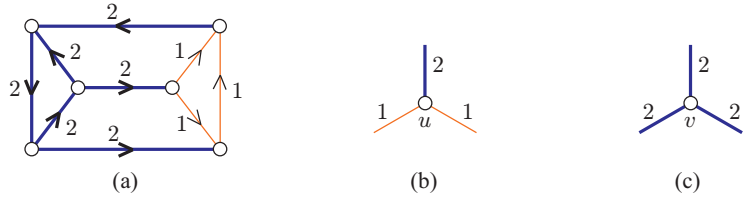}
			\end{center}
			\vspace{-1em}
     		\caption{(a) A configuration of an NCL machine, (b) NCL {\sc and} vertex $u$, and (c) NCL {\sc or} vertex $v$.}
			\label{fig:ncl}
		\end{figure}

	An NCL ``machine'' is an undirected graph together with an assignment of weights from $\{1,2\}$ to each edge of the graph. 
	An ({\em NCL}) {\em configuration} of this machine is an orientation (direction) of the edges such that the sum of weights of in-coming arcs at each vertex is at least two. 
	Figure~\ref{fig:ncl}(a) illustrates a configuration of an NCL machine, where each weight-2 edge is depicted by a thick (blue) line and each weight-1 edge by a thin (orange) line. 
	Then, two NCL configurations are {\em adjacent} if they differ in a single edge direction. 
	Given an NCL machine and its two configurations, it is known to be PSPACE-complete to determine whether there exists a sequence of adjacent NCL configurations which transforms one into the other~\cite{HD05}. 

	An NCL machine is called an {\it {\sc and}/{\sc or} constraint graph} if it consists only of two types of vertices, called ``NCL {\sc and} vertices'' and ``NCL {\sc or} vertices'' defined as follows:
	\begin{enumerate}
	\item A vertex of degree three is called an {\it NCL {\sc and} vertex} if its three incident edges have weights $1$, $1$ and $2$. 
(See \figurename~\ref{fig:ncl}(b).)
	An NCL {\sc and} vertex $u$ behaves as a logical {\sc and}, in the following sense: 
the weight-$2$ edge can be directed outward for $u$ if and only if both two weight-$1$ edges are directed inward for $u$. 
	Note that, however, the weight-$2$ edge is not necessarily directed outward even when both weight-$1$ edges are directed inward. 
\medskip

	\item A vertex of degree three is called an {\it NCL {\sc or} vertex} if its three incident edges have weights $2$, $2$ and $2$. 
(See \figurename~\ref{fig:ncl}(c).)
	An NCL {\sc or} vertex $v$ behaves as a logical {\sc or}: 
one of the three edges can be directed outward for $v$ if and only if at least one of the other two edges is directed inward for $v$. 
	\end{enumerate}
	It should be noted that, although it is natural to think of NCL {\sc and}/{\sc or} vertices as having inputs and outputs, there is nothing enforcing this interpretation; 
especially for NCL {\sc or} vertices, the choice of input and output is entirely arbitrary because an NCL {\sc or} vertex is symmetric. 

	For example, the NCL machine in \figurename~\ref{fig:ncl}(a) is an {\sc and}/{\sc or} constraint graph. 
	From now on, we call an {\sc and}/{\sc or} constraint graph simply an {\it NCL machine}, and call an edge in an NCL machine an {\it NCL edge}. 
	NCL remains PSPACE-complete even if an input NCL machine is planar and bounded bandwidth~\cite{Z15}.

\section{Our Results}

	In this section, we give our results. 
	Observe that {\sc list edge-coloring reconfiguration} can be solved in (most conveniently, nondeterministic~\cite{Sav70}) polynomial space, and hence it is in PSPACE. 
	Therefore, we show the PSPACE-hardness.
	Our reductions from NCL take the same strategy, and hence we first give the common preparation in Section~\ref{subsec:framework}. 
	Then, we prove the PSPACE-hardness of {\sc list edge-coloring reconfiguration} in Section~\ref{sec:03} and the non-list variant in Section~\ref{sec:04}.

\subsection{Preparation for reductions} \label{subsec:framework}

	Suppose that we are given an instance of NCL, that is, an NCL machine and two orientations of the machine. 

\begin{figure}[tb]
		\begin{center}
      		\includegraphics[width=0.85\linewidth]{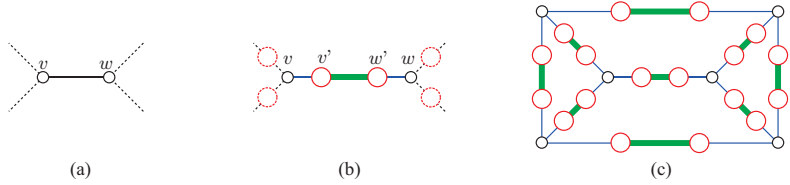}
		\end{center}
		\vspace{-1em}
     		\caption{(a) An NCL edge $vw$, (b) its subdivision into a path $v v^\prime w^\prime w$, and (c) the resulting graph which corresponds to the NCL machine in \figurename~\ref{fig:ncl}(a), where newly added vertices are depicted by (red) large circles, link edges by (green) thick lines and connector edges by (blue) thin lines.}
     		\label{fig:subdivision}
    \end{figure}

	We first subdivide every NCL edge $vw$ into a path $v v^\prime w^\prime w$ of length three by adding two new vertices $v^\prime$ and $w^\prime$.
(See \figurename~\ref{fig:subdivision}(a) and (b).) 
	We call the edge $v^\prime w^\prime$ a {\it link edge} between two NCL vertices $v$ and $w$, and call the edges $vv^\prime$ and $ww^\prime$ {\it connector edges} for $v$ and $w$, respectively. 
	Notice that every vertex in the resulting graph belongs to exactly one of stars $K_{1,3}$ such that the center of each $K_{1,3}$ corresponds to an NCL {\sc and}/{\sc or} vertex.
	Furthermore, these stars are all mutually disjoint, and joined together by link edges.
(See \figurename~\ref{fig:subdivision}(c) as an example.) 

	Our reduction thus involves constructing three types of gadgets which correspond to link edges and stars of NCL {\sc and}/{\sc or} vertices; 
we will replace each of link edges and stars of NCL {\sc and}/{\sc or} vertices with its corresponding gadget. 
	In our reduction, assigning the color $1$ to the connector edge $vv^\prime$ always corresponds to directing the connector edge $vv^\prime$ from $v^\prime$ to $v$ (i.e., the inward direction for $v$), while assigning the color $4$ to $vv^\prime$ always corresponds to directing $vv^\prime$ from $v$ to $v^\prime$ (i.e., the outward direction for $v$).

\subsection{List edge-coloring reconfiguration}\label{sec:03}
		In this subsection, we prove the following theorem.
		\begin{theorem}\label{theorem:List}
		For every integer $k \ge 4$, the {\sc list edge-coloring reconfiguration} problem is PSPACE-complete even if an input graph is planar, of maximum degree three, and has bounded bandwidth.
		\end{theorem}

		\begin{figure}[t]
			\begin{center}
				\includegraphics[width=1.0\textwidth]{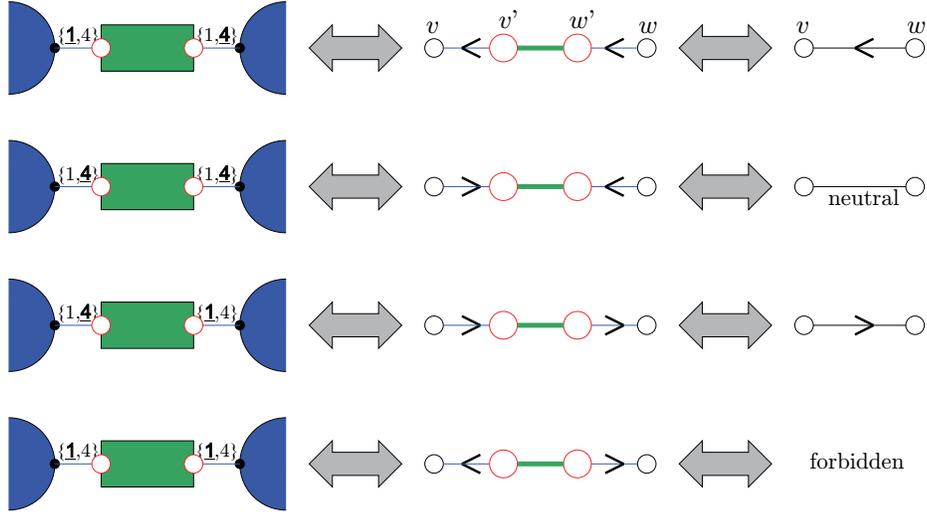}
			\end{center}
			\vspace{-1em}
			\caption{(a) Color assignments to connector edges, (b) their corresponding orientations of the edges $vv^\prime$ and $ww^\prime$, and (c) the corresponding orientations of an NCL edge $vw$.}
			\label{fig:gadgetadj}
		\end{figure}

		We prove the theorem in the remainder of this subsection. 
		As we have mentioned in Section~\ref{subsec:framework}, it suffices to construct three types of gadgets which correspond to link edges and stars of NCL {\sc and}/{\sc or} vertices. 
\medskip

\noindent
			{\bf $\bullet$ Link edge gadget.}

	Recall that, in a given NCL machine, two NCL vertices $v$ and $w$ are joined by a single NCL edge $vw$. 
	Therefore, the link edge gadget between $v$ and $w$ should be consistent with the orientations of the NCL edge $vw$, as follows (see also \figurename~\ref{fig:gadgetadj}):
	If we assign the color $1$ to the connector edge $vv^\prime$ (i.e., the inward direction for $v$), then $ww^\prime$ must be colored with $4$ (i.e., the outward direction for $w$); 
conversely, $vv^\prime$ must be colored with $4$ if we assign $1$ to $ww^\prime$. 
	In particular, the gadget must forbid a list edge-coloring which assigns $1$ to both $vv^\prime$ and $ww^\prime$ (i.e., the inward directions for both $v$ and $w$), because such a coloring corresponds to the direction which illegally contributes to both $v$ and $w$ at the same time. 
	On the other hand, assigning $4$ to both $vv^\prime$ and $ww^\prime$ (i.e., the outward directions for both $v$ and $w$) corresponds to the {\it neutral} orientation of the NCL edge $vw$ which contributes to neither $v$ nor $w$, and hence we simply do not care such an orientation. 

\begin{figure}[t]
				\begin{center}
					\includegraphics[width=0.35\textwidth]{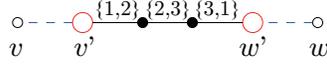}
				\end{center}
				\vspace{-1em}
				\caption{The link edge gadget for {\sc list edge-coloring reconfiguration}.}
				\label{fig:gadgete}
			\end{figure}

\begin{figure}[t]
				\begin{center}
					\includegraphics[width=0.6\textwidth]{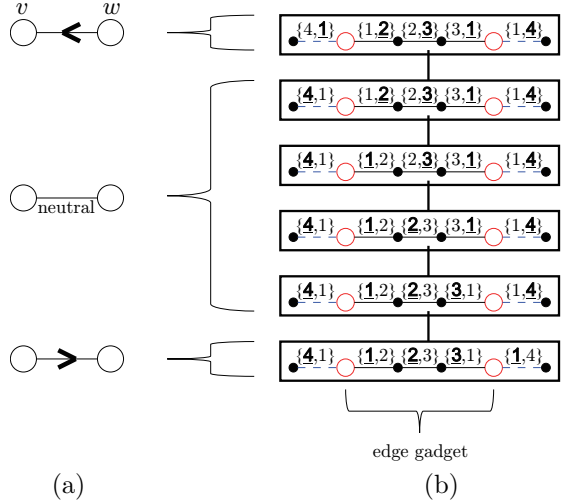}
					\\
					(a)\hspace{45mm}(b)\hspace{8mm}~
				\end{center}
				\vspace{-1em}
				\caption{(a) Three orientations of an NCL edge $vw$, and 
					(b) all list edge-colorings of the link edge gadget with two connector edges.}
				\label{fig:gadgetcorre}
			\end{figure}

			Figure~\ref{fig:gadgete} illustrates our link edge gadget between two NCL vertices $v$ and $w$.
			Figure~\ref{fig:gadgetcorre}(b) illustrates the ``reconfiguration graph'' of this link edge gadget together with two connector edges $vv^\prime$ and $ww^\prime$:
each rectangle represents a node of the reconfiguration graph, that is, a list edge-coloring of the gadget, where the underlined bold number represents the color assigned to the edge, 
and two rectangles are joined by an edge in the reconfiguration graph if their corresponding list edge-colorings are adjacent.
			Then, the reconfiguration graph is connected as illustrated in \figurename~\ref{fig:gadgetcorre}(b), and the link edge gadget has no list edge-coloring which assigns $4$ to the two connector edges $vv^\prime$ and $ww^\prime$ at the same time,
as required.
			Furthermore, the reversal of the NCL edge $vw$ can be simulated by the path
			via the neutral orientation of $vw$, as illustrated in \figurename~\ref{fig:gadgetcorre}(a).
			Thus, this link edge gadget works correctly.

\medskip

\noindent
			{\bf $\bullet$ {\sc And} gadget.}

\begin{figure}[t]
					\begin{center}
						 \includegraphics[width=0.9\textwidth]{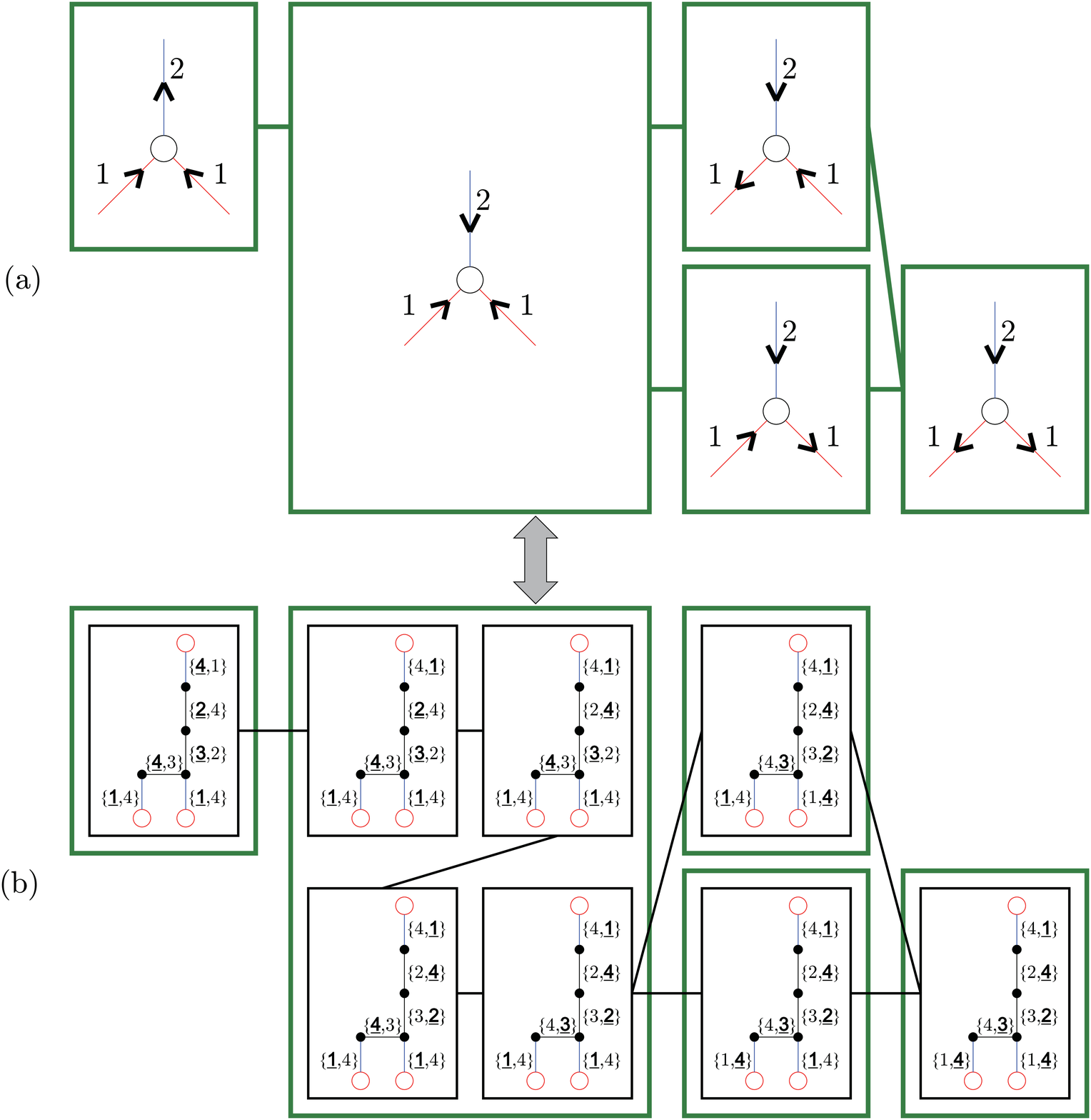}
					\end{center}
				\vspace{-1em}
				\caption{(a) All valid orientations of the three connector edges for an NCL {\sc and} vertex $v$, and (b) 
					all list edge-colorings of the {\sc and} gadget in \figurename~\ref{fig:gadgeta}.}
				\label{fig:gadgetcorra}
			\end{figure}

			Consider an NCL {\sc and} vertex $v$. 
			Figure~\ref{fig:gadgetcorra}(a) illustrates all valid orientations of the three connector edges for $v$;
each box represents a valid orientation of the three connector edges for $v$, and two boxes are joined by an edge if their orientations are adjacent.
			We construct our {\sc and} gadget so that it correctly simulates this reconfiguration graph in \figurename~\ref{fig:gadgetcorra}(a).
			
\begin{figure}[t]
				\begin{center}
					\includegraphics[width=0.15\textwidth]{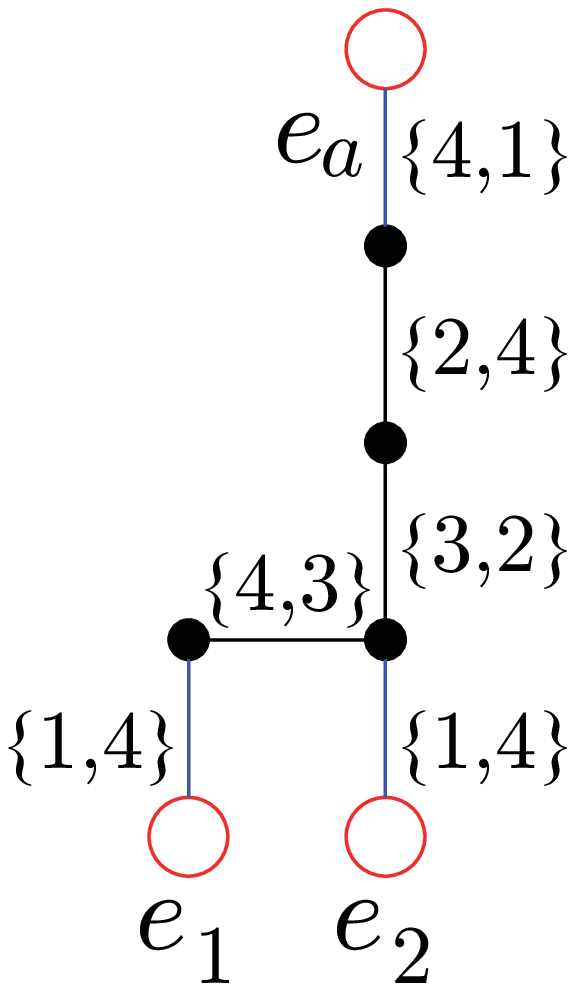}
				\end{center}
				\vspace{-1em}
				\caption{The {\sc and} gadget for {\sc list edge-coloring reconfiguration}.}
				\label{fig:gadgeta}
			\end{figure}

			Figure~\ref{fig:gadgeta} illustrates our {\sc and} gadget for each NCL {\sc and} vertex $v$,
			where $e_1$, $e_2$ and $e_a$ correspond to the three connector edges for $v$ such that $e_1$ and $e_2$ come from the two weight-$1$ NCL edges and $e_a$ comes from the weight-$2$ NCL edge.
			Figure~\ref{fig:gadgetcorra}(b) illustrates the reconfiguration graph for all list edge-colorings of the {\sc and} gadget, where each large box surrounds all colorings having the same color assignments to the three connector edges for $v$. 
			Then, we can see that these
			list edge-colorings are ``internally connected,'' that is, any two list edge-colorings in the same box are reconfigurable with
			each other without recoloring any connector edge.
			Furthermore,
			this gadget preserves the ``external adjacency'' in the following sense: if we contract the list
			edge-colorings in the same box in \figurename~\ref{fig:gadgetcorra}(b) into a single
			vertex, then the resulting graph is exactly the graph depicted in \figurename~\ref{fig:gadgetcorra}(a).
			Therefore, we can conclude that our {\sc and} gadget correctly works as an NCL {\sc and} vertex.

\medskip

\noindent
			{\bf $\bullet$ {\sc Or} gadget.}

			\begin{figure}[t]
				\begin{center}
					\includegraphics[width=0.7\textwidth]{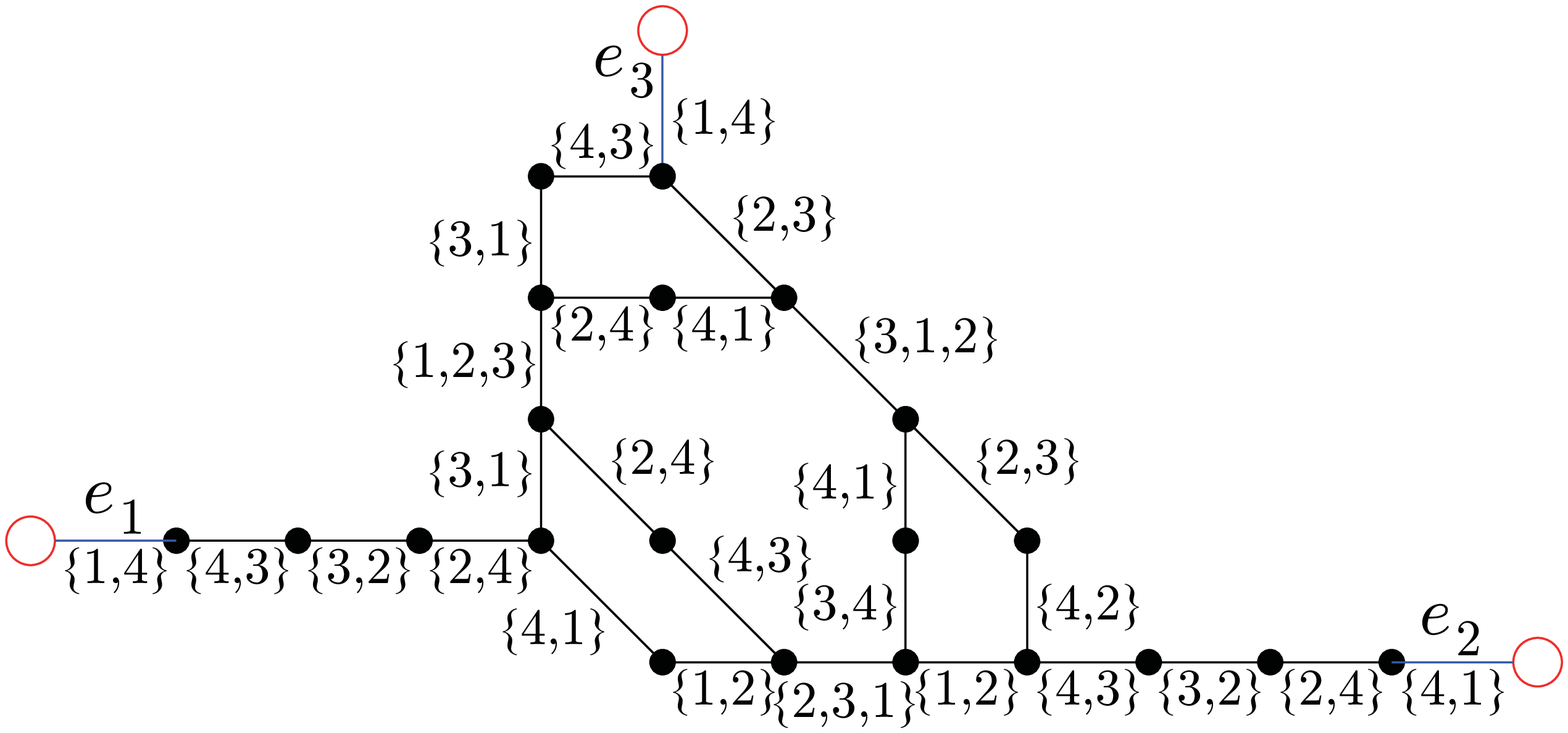}
				\end{center}
				\vspace{-1em}
				\caption{The {\sc or} gadget for {\sc list edge-coloring reconfiguration}.}
				\label{fig:gadgeto}
			\end{figure}

			Figure~\ref{fig:gadgeto} illustrates our {\sc or} gadget for each NCL {\sc or} vertex $v$,
			where $e_1$, $e_2$ and $e_3$ correspond to the three connector edges for $v$.
			To verify that this {\sc or} gadget correctly simulates an NCL {\sc or} vertex, it suffices to show that this gadget satisfies both the internal connectedness and the external adjacency. 
			However, since this gadget has $1575$ list edge-colorings, we have checked that our {\sc or} gadget satisfies these sufficient conditions
			by a computer search of all list edge-colorings of the gadget.
\medskip

\noindent
			{\bf Reduction.}
		
			As we have explained before, we replace each of link edges and stars of NCL {\sc and}/{\sc or} vertices with its corresponding gadget; 
let $G$ be the resulting graph.
			Since NCL remains PSPACE-complete even if an input NCL machine is planar, bounded bandwidth and of maximum degree three, the resulting graph $G$ is also planar, bounded bandwidth and of maximum degree three;
notice that, since each gadget consists of only a constant number of edges, the bandwidth of $G$ is also bounded. 

			In addition, we construct two list edge-colorings of $G$ which correspond to two given
			NCL configurations $C_0$ and $C_r$ of the NCL machine. 
			Note that there are (in general, exponentially)
			many list edge-colorings which correspond to the same NCL configuration. 
			However, by the construction of the three gadgets, no two distinct NCL configurations correspond to the same
			list edge-coloring of $G$. 
			We thus arbitrarily choose two list edge-colorings $f_0$ and $f_r$ of $G$ which correspond
			to $C_0$ and $C_r$, respectively. 

			Clearly, the construction can be done in polynomial time.
\medskip

\noindent
			{\bf Correctness.}

			We now prove that there exists a desired sequence of NCL configurations between $C_0$ and $C_r$ if and only if
			there exists a reconfiguration sequence between $f_0$ and $f_r$.
			
			We first prove the only-if direction.
			Suppose that there exists a desired sequence of
			NCL configurations between $C_0$ and $C_r$, and consider any two adjacent NCL configurations $C_{i-1}$ and $C_i$ in the sequence.
			Then, only one NCL edge $vw$ changes its orientation between $C_{i-1}$ and $C_i$.
			Notice that, since both $C_{i-1}$ and $C_i$ are valid NCL configurations, the NCL {\sc and}/{\sc or} vertices $v$
			and $w$ have enough in-coming arcs even without $vw$.
			Therefore, we can simulate this reversal
			by the reconfiguration sequence of list edge-colorings in \figurename~\ref{fig:gadgetcorre}(b) which passes through the
			neutral orientation of $vw$ as illustrated in \figurename~\ref{fig:gadgetcorre}(a).
			Recall that both {\sc and} and {\sc or} gadgets
			are internally connected, and preserve the external adjacency. Therefore, any reversal of an
			NCL edge can be simulated by a reconfiguration sequence of list edge-colorings of $G$, and hence
			there exists a reconfiguration sequence between $f_0$ and $f_r$.
			
			We now prove the if direction.
			Suppose that there exists a reconfiguration sequence $\langle f_0, f_1, \ldots , f_\ell \rangle$ from $f_0$ to $f_\ell = f_r$.
			Notice that, by the construction of gadgets, any list edge-coloring of $G$
			corresponds to a valid NCL configuration such that some NCL edges may take the neutral orientation.
			In addition, $f_0$ and $f_r$ correspond to valid NCL configurations without any neutral orientation. 
			Pick the first index $i$ in the reconfiguration sequence $\langle f_0, f_1, \ldots , f_\ell \rangle$
			which corresponds to changing the direction of an NCL edge $vw$ to the neutral orientation.
			Then, since the neutral orientation contributes to neither $v$ nor $w$, we can simply ignore the change of the NCL edge $vw$ and keep the direction of $vw$ as the same as the previous direction.  
			By repeating this process and deleting redundant orientations if needed, we can obtain a sequence of valid adjacent orientations between $C_0$ and $C_r$ such that no NCL edge takes the neutral orientation. 

			This completes the proof of Theorem~\ref{theorem:List}.

\subsection{Edge-coloring reconfiguration}\label{sec:04}
		In this subsection, we prove the following theorem for the non-list variant.
		\begin{theorem}\label{theorem:NonList}
		For every integer $k \ge 5$,
		the {\sc edge-coloring reconfiguration} problem is PSPACE-complete even if an input graph is planar whose maximum degree is $k$ and bandwidth is linear in $k$.
		\end{theorem}
			
\begin{figure}[t]
				\begin{center}
					\includegraphics[width=0.5\textwidth]{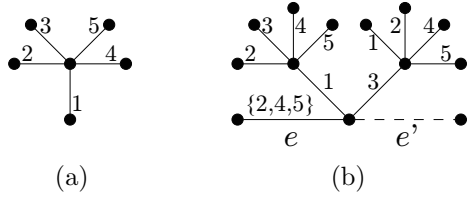}
					\\
					(a) \hspace{31mm}(b) \hspace{6mm}~
				\end{center}
				\vspace{-1em}
				\caption{(a) Gadget for restricting a color ($k=5$), and (b) edge $e$ whose available colors are restricted to $\{2,4,5\}$.}
				\label{fig:pseudo}
			\end{figure}

		To prove the theorem, similarly as in the previous subsection, it suffices to construct three types of gadgets corresponding to a link edge and stars of NCL {\sc and}/{\sc or} vertices. 
		However, since we deal with the non-list variant, every edge has all $k$ colors as its available colors. 
		Thus, we construct one more gadget, called a {\em color gadget}, which restricts the colors available for the edge. 
		The gadget is simply a star having $k$ leaves, and we assign the $k$ colors to the edges of the star in both $f_0$ and $f_r$.
(See \figurename~\ref{fig:pseudo}(a) as an example for $k=5$.) 
		Note that, since the color set consists of only $k$ colors, these $k$ edges must stay the same colors in any reconfiguration sequence. 
		Thus, if we do not want to assign a color $c$ to an edge $e$, then we connect the leaf with the color $c$ to an endpoint $v$ of $e$. 
(See \figurename~\ref{fig:pseudo}(b).) 
		In this way, we can treat the edge $e$ as if it has the list $L(e)$ of available colors. 
		However, we need to pay attention to the fact that all edges $e^\prime$ sharing the endpoint $v$ cannot receive the color $c$ by connecting such gadgets to $v$. 

			\begin{figure}[t]
				\begin{center}
					\includegraphics[width=0.7\textwidth]{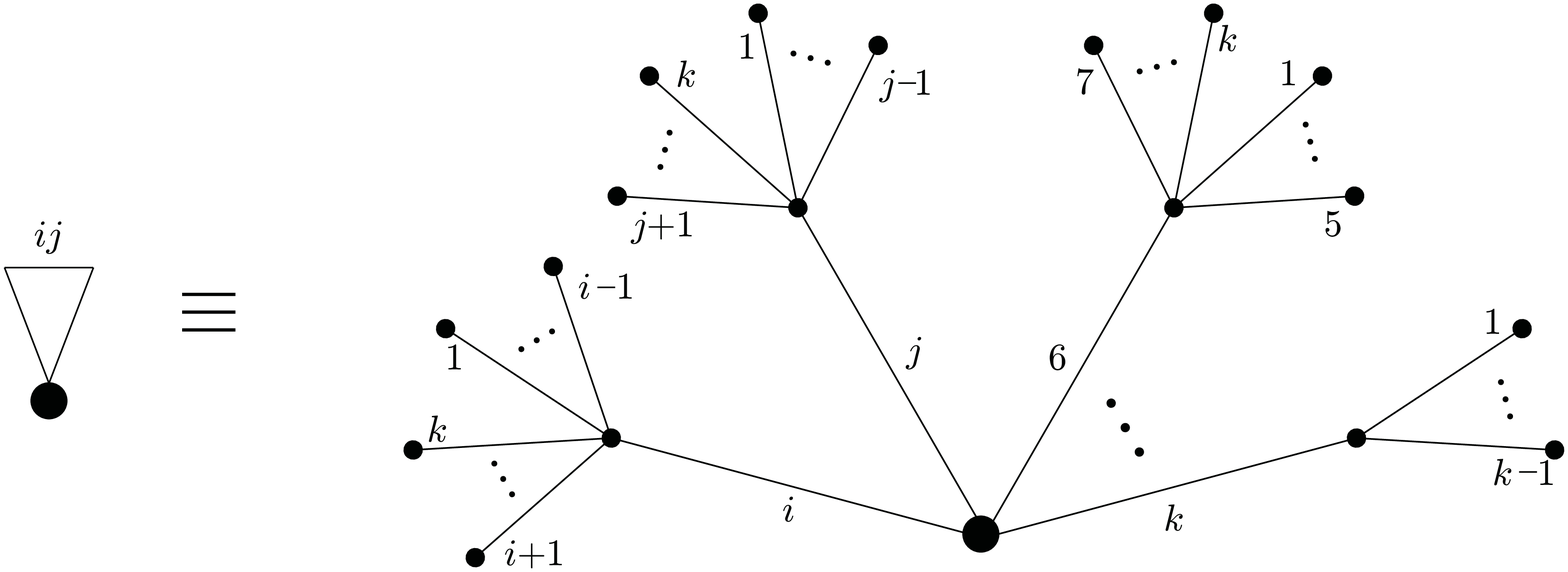}
				\end{center}
				\vspace{-1em}
				\caption{Explanatory note for the gadgets in \figurename~\ref{fig:allgadget}.}
				\label{fig:explanatory}
			\end{figure}

			\begin{figure}[t]
				\begin{center}
					\includegraphics[width=1.0\textwidth]{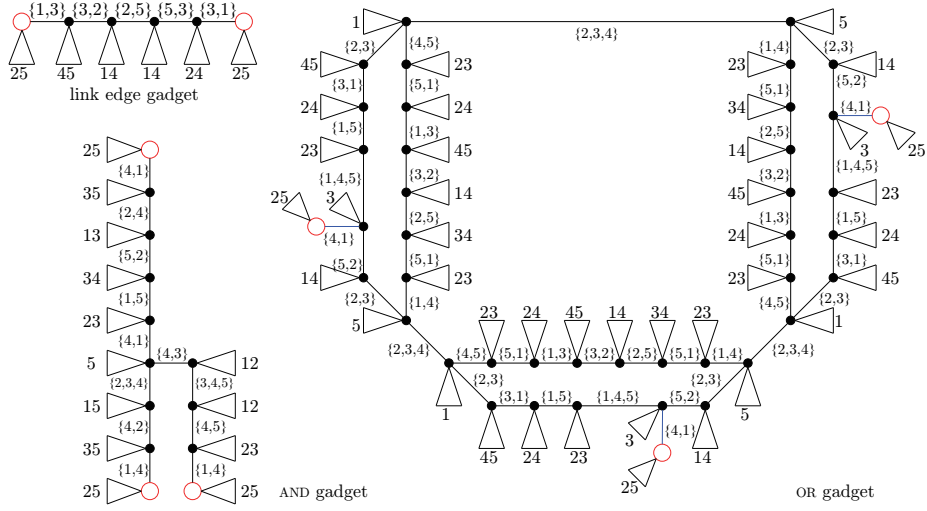}
				\end{center}
				\vspace{-1em}
				\caption{The link edge, {\sc and}, and {\sc or} gadgets for the non-list variant.}
				\label{fig:allgadget}
			\end{figure}

			Figures~\ref{fig:explanatory} and \ref{fig:allgadget} illustrate all gadgets for the non-list variant, where the available colors for each edge is attached as the list of the edge.  
			Notice that the gadget in \figurename~\ref{fig:explanatory} forbids the colors $i$, $j$, and $6, 7, \ldots, k$. 
			We emphasize that all (red) endpoints of connector edges shared by other gadgets are attached with the same color gadgets which forbid two colors $2$ and $5$, and hence we can connect the gadgets consistently. 
			Then, the link edge gadget and {\sc and}/{\sc or} gadgets have $6$, $40$, $477192$ edge-colorings, respectively.
			We have checked that all gadgets satisfy both the internal connectedness and the external adjacency 
			by a computer search of all edge-colorings of the gadgets.
			Therefore, by the same arguments as the proof of Theorem~\ref{theorem:List}, we can conclude that an instance of NCL is a $\sf{yes}$-instance if and only if the corresponding instance of {\sc edge-coloring reconfiguration} is a $\sf{yes}$-instance.

			Recall that NCL remains PSPACE-complete even if an input NCL machine is planar, bounded bandwidth, and of maximum degree three. 
			Thus, the resulting graph $G$ is also planar. 
			Notice that only the size of the color gadget depends on $k$, and the other (parts of) gadgets are of constant sizes. 
			Since $k \ge 5$, the maximum degree of $G$ is $k$, the degree of the center of each color gadget. 
			In addition, since each of link edge and {\sc and}/{\sc or} gadgets contains only a constant number of color gadgets, the number of edges in each gadget can be bounded by a linear in $k$. 
			Since the bandwidth of the input NCL machine is a constant, that of $G$ can be bounded by a linear in $k$.  
			
			This completes the proof of Theorem~\ref{theorem:NonList}.

\section{Conclusion}\label{sec:05}

	In this paper, we have shown the PSPACE-completeness of {\sc list edge-coloring reconfiguration} and its non-list variant. 
	We emphasize again that our result for {\sc list edge-coloring reconfiguration} gives a sharp analysis of the complexity status with respect to the number $k$ of colors. 
	In addition, our result is the first complexity hardness result for the non-list variant. 


\end{document}